\documentclass[fleqn,usenatbib]{mnras}

\usepackage{newtxtext,newtxmath}

\usepackage[T1]{fontenc}

\DeclareRobustCommand{\VAN}[3]{#2}
\let\VANthebibliography\thebibliography
\def\thebibliography{\DeclareRobustCommand{\VAN}[3]{##3}\VANthebibliography}
\DeclareSymbolFont{matha}{OML}{txmi}{m}{it}
\DeclareMathSymbol{\varv}{\mathord}{matha}{118}

\usepackage{graphicx}	
\usepackage{amsmath}	
\usepackage{caption}
\usepackage{subcaption}

\title[Waves in multi-fluid weakly ionised solar plasmas]{Waves in weakly ionised solar plasmas}

\author[A. Alharbi et al.]{
A. Alharbi,$^{1,2}$\thanks{E-mail: aalharbi8@sheffield.ac.uk; ahhrbe@uqu.edu.sa}, 
I. Ballai,$^{1}$
V. Fedun$^{3}$
and G. Verth$^{1}$
\\
$^{1}$Plasma Dynamics Group, School of Mathematics and Statistics, University of Sheffield, Hounsfield Road, Hicks Building, Sheffield, S3 7RH, UK\\
$^{2}$Department of Mathematics, Jamoum University College, Umm Al-Qura University, Jamoum, 25375 Makkah, Saudi Arabia\\
$^{3}$Plasma Dynamics Group, Department of Automatic Control and Systems Engineering, The University of Sheffield, Mappin Street, Sheffield, S1 3JD, UK
}

\date{Accepted 2022 February 14 . Received 2022 February 11; in original form 2022 January 13}

\pubyear{2021}

\begin{document}
\label{firstpage}
\pagerange{\pageref{firstpage}--\pageref{lastpage}}
\maketitle

\begin{abstract}
Here we study the nature and characteristics of waves propagating in partially ionised plasmas in the weakly ionised limit, typical for the lower part of the solar atmosphere. The framework in which the properties of waves are discussed depends on the relative magnitude of collisions between particles, but also on the relative magnitude of the collisional frequencies compared to the gyro-frequency of charged particles. Our investigation shows that the weakly ionised solar atmospheric plasma can divided into two regions and this division occurs, roughly, at the base of the chromosphere. In the solar photosphere the plasma is non-magnetised and the dynamics can described within the three-fluid framework where acoustic waves associated to each species can propagate. Due to the very high concentration of neutrals, the neutral sound waves propagates with no damping, while for the other two modes the damping rate is determined by collisions with neutrals. The ion and electron-related acoustic modes propagate with a cut-off determined by the collisional frequency of these species with neutrals. In the weakly ionised chromosphere only electrons are magnetised, however, the strong coupling of charged particles reduces the working framework to a two-fluid model. The disassociation of charged particles creates electric currents that can influence the characteristic of waves. The propagation properties of waves with respect to the angle of propagation are studied with the help of polar diagrams.  
\end{abstract}




\section{Introduction}

Study of waves and oscillations in the solar atmosphere received new impetus in the last few decades thanks to a plethora of high-resolution observations showing plasma dynamics on all sort of spatial and temporal scales. Waves proved to be an essential tool for plasma and magnetic field diagnostics when combining theoretical results (dispersion relations, evolutionary equation, polarization of waves, etc.) with observations (wavelengths, periods, damping time and length, etc.). 

In recent years there was a substantial surge in the number of studies of waves where the plasma was considered to be partially ionised, describing a situation one can meet in the lower part of the solar atmosphere where the temperature is not high enough to ensure a complete ionisation of the plasma. In such plasmas the governing equations describing the dynamical and thermodynamical  state of the plasma requires a multi-fluid approach, where particles interact via collisions and the frequencies at which changes occur are comparable to the collisional frequencies of particles (Zaqarashvili et al. 2011, Khomenko et al. 2014). A multi-fluid approach not only increases considerably the number of equations (and so the number of waves present in such plasmas) and the complexity of the mathematical description, but also involves some physical aspects that cannot be recovered in a single fluid approximation (e.g. existence of cut-offs, forbidden propagation regimes, plasma filamentation, etc.). 

The study of waves and instabilities in partially ionised plasmas has shed light on the effect of interaction between particles on the nature and properties of waves, and on the onset and evolution of instabilities. In this respect Zaqarashvili et al. (2011), Soler et al. (2013b) and Popescu-Braileanu et al. (2019) studied the propagation of magnetoacoustic waves in two-fluid plasma, where the two components are the charged particles (positive ions and electrons) and neutrals. Their results showed that in this configuration waves are affected by the collision between particles, leading to a damping of waves. In addition, when considering chromospheric conditions, these studies found that magnetoacoustic waves with wavelengths smaller than 1 km are affected by two-fluid effects in regions with intense magnetic fields, while much shorter wavelengths have to be considered for these effects to be relevant in quiet Sun conditions. 

The same two-fluid model has been also considered when analysing Alfv\'en waves by Soler et al. (2013a). These authors found that similar to magnetoacoustic modes, Alfv\'en waves are also strongly influenced by the collision between particles and the damping of these modes is most efficient when the wave frequency and the collision frequency are of the same order of magnitude. The effect of heavy particles (He atoms) and stratification on the propagation of torsional Alfv\'en waves was investigated by Zaqarashvili et al. (2013) and their results show that shorter-period (< 5 s) torsional Alfv\'en waves damp quickly
in the chromospheric network due to ion-neutral collision, while longer-period (> 5 s) waves do not reach the transition region as they become evanescent at lower heights in the network cores, meaning that stratification of the plasma has a filtering effect on wave propagation. 

The partially ionised solar plasma in the lower part of the solar atmosphere is a very dynamical environment, where the chemical composition of the plasma can change over time scales comparable with the temporal characteristic of waves due to additional ionisation and recombination. These processes confer the plasma a non-equilibrium state. Waves in such plasmas have been modelled numerically and analytically (e.g. Maneva et al. 2017, Ballai 2019, Popescu-Braileanu 2019, Zhang et al. 2021) showing that the propagation of waves and the process of plasma heating is seriously affected by non-equilibrium effects. For a comprehensive review on the property of waves in partially ionised plasma see, e.g. Ballester et al. (2018).   

In a recent study, Alharbi et al. (2021) studied the propagation of guided slow sausage waves in the presence of gravitational stratification in the solar chromosphere, where the ionisation degree is high. They used a two-fluid model (neutrals and charged particles), and the small value of the density ratio between neutrals and ions was employed as an expansion parameter. Considering an initial value problem, their analysis showed that while ion-acoustic modes possessed an acoustic cut-off, the slow mode associated with neutrals propagated with no cut-off frequency and the absence of this value was due to the collisions between neutrals and ions. 

The above studies considered a two-fluid approach, however, their model choice was not always justified within the framework of solar and space plasmas. In many investigations listed earlier the collisional frequency was considered a free parameter and the analysis focused on the investigation of wave properties when the collisional frequency was varying between the collision-free case and a regime completely dominated by collisions. While such investigations are important to analyse the variation of wave properties over a large parametric space, these have a rather restricted connectivity to real solar physics applications. In the present study we plan to apply observational constrains and construct our model on realistic background. 

Our research can be considered to be a continuation (or complementary) of the study described in Alharbi et al. (2021), but now we focus on the study of waves in weakly ionised limit, corresponding to the lower part of the solar atmosphere. 

We will start by constructing our working environment using standard solar atmospheric models and investigate the key properties that are important for our purposes (Section 2). Section 3 is devoted to the introduction of mathematical formalism together with assumptions that are based on observational facts aimed to simplify our treatment as much as possible, yet describing a realistic physical situation. The governing equations describing the evolution of waves are solved and analysed in Section 4 for a simple configuration to elucidate the role of gravitational stratification in the process of wave propagation. A normal mode analysis is employed in for different frequency regimes in order to study the properties of waves and their propagation characteristics. Finally, our conclusions are presented in Section 5. 

\section{Model restrictions}

In order to correctly describe the evolution of waves and their properties, we will need to impose a few physical restrictions derived from observations. The lower part of the solar atmosphere is a layer of the solar atmosphere which is characterised by relatively low temperatures and high densities. This means that here the plasma is weakly ionised. Various solar atmospheric models predict a ratio of neutral number density to ions of the order $10^4$ (for details see Alharbi et al. 2021). Since we assume a quasi-neutral hydrogen plasma, the above statement also means that the ratio between the number density of neutrals and electrons is equally very high. 

During their motion particles interact through short and long-range collisions, during which the plasma can be thermalised and effective momentum transfer can take place. While electrons have much lower mass than ions and neutrals and they do not contribute significantly to the process of momentum and energy transfer, however, their interaction with massive particles (neutrals and ions) are important because any thermal inhomogeneity will be smoothed out through the electron collisions. Given the high density of particles in the solar photosphere, collisions will play a very important role.

Another important ingredient in our analysis is the magnetic field. The region of our interest corresponding to low level of ionisation, is the region whose magnetic field, its properties and structure, was determined and measured most reliably using various techniques (e.g. spectroscopic measurements, the Zeeman-and Hanle-effects, etc.). The magnetic field in the solar photosphere takes various forms. The most prominent structures dominating the active regions and the network are the sunspots and pores (appearing as dark regions), that have field strengths of a few kG. In addition, the quiet Sun is permeated by a weaker field, often forming structures such as small-scale $\Omega$ loops, $U$-loops and turbulent field. Although such features are found everywhere in the photosphere, they are most typical for the quiet Sun, in particular the interiors of supergeranular cells, the internetwork (for a review of the quiet Sun magnetism see, e.g. Sanchez Almeida 2003, Khomenko 2006, Trujillo Bueno et al. 2006, Title 2007, Solanki 2009). In our study we will focus on weaker field regions and assume a simple variation of the magnetic field with height. 

The formulaic variation of the magnetic field with height is an open question and this depends very much on the nature of the magnetic structure we would like to investigate. 
Applying the thin flux-tube model and pressure balance equation Khomenko et al. (2015) proposed a magnetic field variation law with height of the form
\begin{equation}
B_0(z)=B_0\exp\left[-\frac{z}{600}\right], 
\label{eq:1}
\end{equation}
where $B_0$ is the magnetic field at $z=0$ reference level and the height, $z$, is measured in km. We should mention here that the choice of the above dependence is somehow arbitrary. Using the same considerations Vranjes and Krstic (2013) proposed a height variation where the e-folding length was 250 km. The same thin flux tube principle was used by Goodman (2000) who, inspired by a series of considerations made by Parker (1979), assumed a height dependence of the magnetic field of the form
\[
B(z)=B_0 \exp[-\Gamma(z)],
\]
where 
\[
\Gamma(z)=\frac{m_i g}{2k_B}\int_0^z\frac{dz'}{T(z')},
\]
with $m_i$ being the ion mass, $g=274$ m s$^{-2}$ the gravitational acceleration, $k_B$ the Boltzmann constant, and $T(z)$ the height-dependent temperature. Finally, Martinez-Pillet et al. (1997) proposed a height variation of the magnetic field above the plage region of the form 
\[
B(z)=B_0\left[\frac{\rho(z)}{\rho(0)}\right]^a,
\]
where the value of the constant $a$ is taken to be 0.3 and $\rho(0)$ is the density at height $z=0$. While the first two relations give an empirical dependence on the height, the last two equations connect the variation of the magnetic field with height to the variation of the temperature and density, respectively. In our study we are going to use the relation proposed by Khomenko et al. (2015) assuming a 100 G magnetic field at the $z=0$ level. At the $z=2$ Mm chromospheric height this model predicts a magnetic field of 3.57 G. 

With the values of the magnetic field determined using equation (\ref{eq:1}) we can determine the values of the electron and ion gyro (or cyclotron) frequencies, defined as
\[
\Omega_{Be}=\frac{eB}{m_e}, \quad \Omega_{Bi}=\frac{eB}{m_i},
\]
with $\Omega_{Be}=\mu\Omega_{Bi}$, where $\mu$ is the ion-to-electron mass ratio. For a hydrogen plasma this ratio is $\mu=m_i/m_e\approx 1836$.

\begin{figure}
   	\includegraphics[width=\columnwidth]{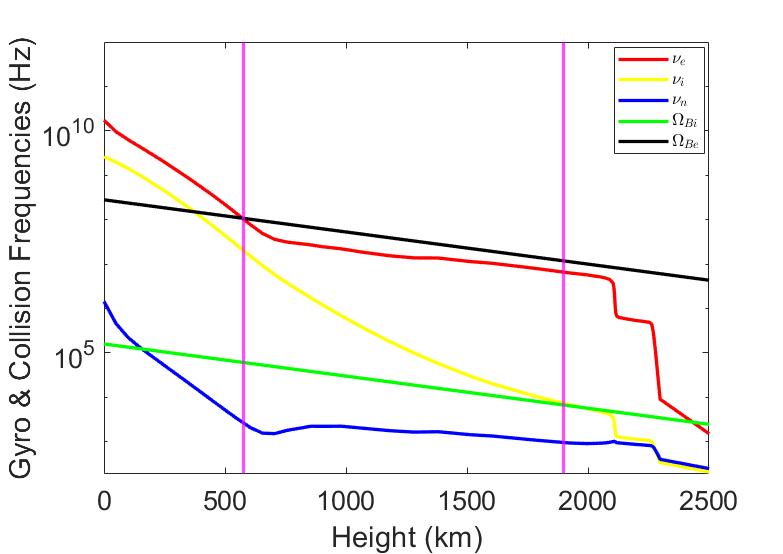}
    \caption{The variation of the collisional frequency of various particles and gyro-frequencies with height based on the VAL III atmospheric model (Vernazza et al. 1981). The pink vertical lines are showing the locations where the collisional frequencies of electrons and ions cross the electron and ion gyro-frequencies. The vertical lines delimitate   regions in the solar atmosphere with different dynamics.  }    \label{frequencies}
   \end{figure}

The cyclotron frequencies are the frequencies at which the charged particles gyrate around the magnetic field lines and these are going to play an important role in the determination of the nature of the dynamics. The variation of these two important quantities with height is shown in Figure \ref{frequencies} by solid green and black lines, respectively (on logarithmic scale). 
It is clear that in the case of both species the gyro-frequency decays with height.

The second key physical ingredient of our problem is the collision between particles. While collisions between neutral and charged species is a short-range collision (head-on), the collision between charged particles, is a long-range collision, controlled by electrostatic forces.  

In order to quantify the importance of collisions, we make use of the solar atmospheric model developed by Vernazza et al. (1981), labelled as the VAL III model. This model describes the variation of several physical parameters in the quiet Sun (e.g. temperature, pressure, number density of neutral hydrogen, number density of electrons, etc.) with height based on spectroscopic measurements. The height described by this model covers the region from the base of the photosphere to the upper chromosphere. 

The collisions between various particles are important as these processes ensure an effective transfer of energy and momentum between species, and provide a mechanism for thermalisation of the plasma.

Let us assume that the collision between particles are elastic and no further ionisation/recombination, excitation and charge exchange take place. Considering species having the same temperature, the binary collisional frequencies between species can be given as (for a detailed description of transport processes in partially ionised plasmas, see, e.g. Zhdanov 1962) 
\begin{equation}
    \nu_{\alpha\beta}=\frac43n_{\beta}\left(\frac{8k_BT}{\pi m_{\alpha\beta}}\right)^{1/2}\sigma_{\alpha\beta},
    \label{eq:2.2}
\end{equation}
where $n_\beta$ is the number density of species $\beta$, $\sigma_{\alpha\beta}$ is the collisional cross-section of the two species and 
\[
m_{\alpha\beta}=\frac{m_{\alpha}m_{\beta}}{m_{\alpha}+m_{\beta}},
\]
is the reduced mass of the system forming particles of mass $m_{\alpha}$ and $m_{\beta}$. Since we are considering only elastic collisions when the rate of momentum transfer is conserved, 
\[
m_{\alpha}n_{\alpha}\nu_{\alpha\beta}=m_{\beta}n_{\beta}\nu_{\beta\alpha}.
\]

For the collision between charged particles an electron will be affected
by a neighbouring ion if the Coulomb potential is of the order (or more than) the electron thermal energy $3k_BT/2$.
This defines a Coulomb interaction distance, $r_C$
\[
\frac{e^2}{4\pi\epsilon_0r_C}\approx \frac32 k_BT,
\]
where $e$ is the electron charge, and $\epsilon_0$ is the permittivity of the vacuum. With the help of this distance (often called the distance of the closest approach) we can define the collisional cross-section of collision between electrons and ions as
\[
\sigma_{ei}=\pi r_C^2\ln \Lambda \approx \pi\left(\frac{e^2}{6\pi\epsilon_0}\right)^2\frac{1}{k_B^2T^2}\ln \Lambda,
\]
where
\[
\ln \Lambda=23.4-1.16\log_{10}n_e+3.45\log_{10} T,
\]
is the Coulomb logarithm, with $n_e$ measured in cm$^{-3}$ and $T$ in K. The Coulomb logarithm is needed in the expression of the collisional cross section to account for the truncation related to small angle scattering. Under solar atmospheric conditions the Coulomb logarithm takes a value in the range 10-25.

With the help of this cross section the collisional frequency between electrons and ions becomes
\begin{equation}
\nu_{ei}=\frac43n_i\sigma_{ei}\left(\frac{8k_BT}{\pi m_e}\right)^{1/2}.
\label{eq:2.3}
\end{equation}

For the collision interaction between ions and neutrals, the collisonal frequency is given by
\begin{equation}
\nu_{in}=\frac{16}{3}n_n\sigma_{in}\left(\frac{k_BT}{\pi m_i}\right)^{1/2},
\label{eq:2.4}
\end{equation}
where $\sigma_{in}$ is the collisional cross-section between ions and neutrals. Based on the study by Vranjes and Krstic (2013), we consider the value of $\sigma_{in}= 10^{-18}$ m$^2$ as representative for photoposheric conditions. In the above relation $n_n$ is the number density of neutrals. 

Finally, the collision between electrons and neutrals can be given as
\begin{equation}
\nu_{en}=\frac43n_n\sigma_{en}\left(\frac{8k_BT}{\pi m_e}\right)^{1/2}.
\label{eq:2.5}
\end{equation}
Assuming a solid sphere collision between electrons and neutrals, the collisional cross-section takes the value $\sigma_{en}=10^{-19}$ m$^2$ (Draine et al. 1983) 

Clearly, due to the height variation of the number densities and temperatures, the collisional frequencies between the three species will also vary with height. To illustrate this we employ the VAL III model (Vernazza et al. 1981) and we plot the total collisional frequencies of particles together with the gyro-frequencies of charged particles with height (see Figure 1). In this figure the total collisional frequency denote the collisional frequency of a particular species with the other two species, e.g. $\nu_i=\nu_{ie}+\nu_{in}$. The lines corresponding to the electron, ion and neutral total collisional frequencies are shown by red, yellow and blue solid lines, respectively.

Figure \ref{frequencies} constitutes the physical background of the investigation presented here. We can clearly identify several regimes will distinct physics governing the evolution of the plasma. First of all below heights of approximately 600 km, (Region I) covering the extent of the whole photosphere, the collisional frequencies of both charged species are larger than the gyro-frequencies of both electrons and ions, meaning that in this region particles collide many times within a gyro-period, so magnetic effects can be neglected. In the absence of magnetic forces particles have a simple Brownian motion and the dynamics can be described within the framework of usual hydrodynamics. Above this height (Region II), up to approximately 1900 km, both the collisional frequency of charged species become smaller than the electron gyro-frequency, but they are still larger than the ion gyro frequency, meaning that electrons become magnetised, however, ions are still non-magnetised and they are collisional coupled to neutrals. That implies that magnetic forces affect only the motion of electrons. Finally, above the height of 1900 km the collisional frequency of both charged species becomes smaller than the ion gyro-frequency, meaning that ions also become magnetised. For completeness, electrons remains unmagnetised as long as the magnetic field is below the critical value
\[
B_c<\frac{2\pi \nu_{en}m_e}{e}.
\]
Considering a characteristic electron-neutrals collisional frequency of $5\times 10^9$~Hz, the critical magnetic field takes the value of 5.5~kG. Since such magnetic fields are unrealistically high in the quiet solar photosphere, our working framework is valid for any magnetic field. Sunspots are known to have kG magnetic fields, however, there the atmospheric model employed by us is not applicable. The solar photosphere is the region of the solar atmosphere where one can have not only very intensive magnetic fields (e.g. sunpots, knots, pores are typical examples where field strength is of the order of a few kG), but also a remnant quiet Sun magnetic field of the order of a few tens of G (Ballai and Forgacs-Dajka 2010).  However, the delimitations of the above regions are not unique and depend very much not only on the physical parameters of the plasma (number density, temperature), but also on the basal value of the magnetic field and its dependence on height. Clearly, for a larger magnetic field the interface that separates non-magnetised region and the region where electrons are magnetised moves towards the solar surface. Given the very high value of the critical magnetic field, it is clear that we are going to always have a region where the collisional frequency of particles is above the electron gyro-frequency. Indeed, for a basal magnetic field of 2~kG, as used in the study by Soler et al. (2015), the upper boundary of region I becomes 486 km above the solar surface.

\begin{figure}
   	\includegraphics[width=\columnwidth]{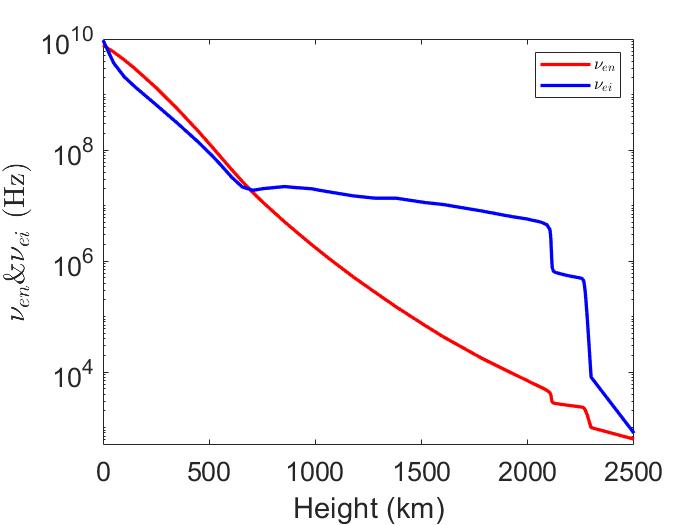}
    \caption{The variation of the collisional frequency of electrons with ions ($\nu_{ei}$, blue line) and electrons with neutrals ($\nu_{en}$, red line) with height based on a VAL IIIC solar atmospheric model (Vernazza et al. 1981).}
         \label{Fig2}
   \end{figure}

Another key aspect that can be derived from variation of collisional frequencies of electrons with height (see Figure \ref{Fig2}) is that starting from the low chromosphere electrons are much more tightly coupled to ions than to neutrals. As a consequence the charged particles can form a single fluid that can interact with the neutral fluid. Below this height, the collisional frequencies of electrons and the other species are  close to each other, meaning that the plasma dynamics can be described within the framework of a three-fluid plasma.   

On the other hand, in Region II electrons are magnetised, however, the collision of electrons with ions and neutrals will not change the dynamics of the system, when the frequency domain is restricted to frequencies of the order of ion-neutral collisional frequency, i.e. the frequency regime described by a two-fluid approximation, where ions and electrons are strongly coupled and form a charged fluid that interacts with the neutral fluid. 

Given the relatively low temperature of the lower solar atmosphere, the gravitational scales heights might be comparable with the wavelength of waves, meaning that gravitational stratification could play an important role in the propagation of waves. This aspect will be elucidated later in this paper. 

In their recent study Alharbi et al. (2021) found that a multi-temperature plasma can reach thermal equilibrium very fast,  and they found that this time is given by
considering the weakly ionised environment discussed in the present paper ($n_n\gg n_i$) in the case of the collision between ions and neutrals, their relation reduces to
\[
t_f=\frac{1}{\nu_{in}}\left(4.6+\ln\left|\frac{{\hat T}_i}{{\hat T}_n}-1\right|\right),
\]
where ${\hat T}_i$ and ${\hat T}_n$ are the temperatures of the ion and neutral fluids and $\nu_{in}$ is the ion-neutral collisional frequency. For an order of magnitude estimate let us consider that ${\hat T}_i=3{\hat T}_n$, and $\nu_{in}=5\times 10^6$ s$^{-1}$. As a result, the time needed for the two species to reach 99\% of the thermal equilibrium is $10^{-8}$ seconds, i.e. thermal equilibrium between the massive particles is settled, indeed, very quickly, much quicker than in the upper part of the atmosphere, thanks to the very high collisional frequency. This conclusion is in line with the results obtained by earlier studies (e.g. Soler et al. 2013 and Oliver et al. 2016). A similar conclusion can be drawn after analysing the time required to reach the equilibrium temperature between electrons and neutrals (where $t_f$ is now of the order of $\nu_{en}^{-1}$). As a result, it is reasonable to consider that the temperature of species equalises very rapidly and, therefore, the assumption of uni-thermal plasma is justified. In a uni-thermal plasma the sound speeds associated to different species will be equal.  

\section{Governing Equations and Assumptions}

The dynamics of waves studied in this paper can be described within the framework of a multi-fluid MHD approximation, where the governing equations describe conservation laws for each species (electrons, ions, neutrals) denoted by the index $\alpha$ ($\alpha=e, i, n$)

\begin{equation}
\frac{\partial \rho_\alpha}{\partial t}+\nabla\cdot (\rho_\alpha{\bf v}_\alpha)=0,
\label{eq:3.1.1}
\end{equation}
\begin{equation}
\rho_\alpha\frac{d{\bf v}_\alpha}{\partial t}+\nabla p_\alpha=\rho_\alpha{\bf g}+\sum_{\alpha^{\prime}\neq \alpha}{\bf P}_{\alpha\alpha^{\prime}}+{\bf \cal F},
\label{eq:3.1.2}
\end{equation}
\begin{equation}
\frac{d}{dt}\left(\frac{p_\alpha}{\rho_\alpha^{\gamma}}\right)=0,
    \label{eq:3.1.3}
\end{equation}
where ${\bf \cal F}$ represents any other force that will be introduced later in Section 4.1, $\rho_\alpha$, ${\bf v}_\alpha$ and $p_\alpha$ are the density, velocity vector and pressure of species $\alpha$, $\gamma$ is the adiabatic index, and ${\bf P}_{\alpha\alpha^{\prime}}$ represents the longitudinal momentum transfer between species $\alpha$ and $\alpha^{\prime}$, and the summation is made for $\alpha^{\prime}\neq \alpha$. We should note that, under normal circumstances, the above energy equations should contain a heating term due to collision between particles, however, since this term is proportional to the square of the velocities (i.e. nonlinear). Since in all subsequent calculations we will limit ourselves to a linear approximation, these terms will be consistently neglected.

\section{Waves in weakly ionised plasmas}

Based on the complex picture revealed by Figure \ref{frequencies}, it is natural to divide the region containing weakly ionised plasma in different regimes and study the properties of waves for each separate case.

\subsection{Waves in Region I}

The spatial extent of this region is determined by the condition $\nu_{e}>\Omega_{Be}$. For the chosen magnetic field profile, this region covers the whole photosphere. The fact that magnetic effects can be neglected when concentrating on the dynamics of waves in weakly ionised photospheric plasma, considerably simplifies the modelling because the only waves that can propagate are the acoustic modes. In the present study we are going to study only linear waves, therefore all governing equations will be linearised. 

Since, in the first instance, we would like to elucidate the importance of gravity and the effects connected to this important ingredient, we are going to concentrate on one-dimensional dynamics and consider that perturbations propagate vertically, against the gravitational field. That is why the dynamics of waves in the three-fluid plasma will be given by the linearised system of equations (\ref{eq:3.1.1})--(\ref{eq:3.1.3}) as 
 \begin{equation} \label{eq3.1}
     \frac{\partial \rho_{\alpha}}{\partial t}+\frac{\partial}{\partial z}(\rho_{0\alpha} v_{\alpha})=0,
 \end{equation}
\begin{equation}\label{eq3.3}
    \rho_{0\alpha} \frac{\partial v_{\alpha}}{\partial t}+ \frac{\partial p_{\alpha}}{\partial z}+ \rho_{\alpha} g=\sum_{\alpha^{\prime}\neq \alpha}P_{\alpha\alpha^{\prime}},
\end{equation}
\begin{equation}  \label{eq3.5}
    \frac{\partial p_{\alpha}}{\partial t}+ v_{\alpha} \frac{dp_{0\alpha}}{dz}=c_{S}^2 \left(\frac{\partial \rho_{\alpha}}{\partial t}+v_{\alpha} \frac{\partial \rho_{\alpha}}{\partial z} \right).
\end{equation}

In the above equations the quantities with an index '0' denote equilibrium values, $c_{S}=\left(\gamma {p_{0\alpha}}/\rho_{0\alpha}\right)^{1/2}$ is the sound speed of species $\alpha$. Since we consider uni-thermal plasma, the sound speeds for all three species will be equal, therefore we will drop the index $\alpha$ from the sound speeds. The quantities $P_{\alpha\alpha^{\prime}}$ refer only to the longitudinal transfer of momentum between species and the velocity $v_{\alpha}$ refers to the vertical component of velocity.

Following the study by Zhdanov (1962), the momentum transfer per unit volume between various species of particles can be written as
\begin{equation}
    P_{\alpha\alpha^{\prime}}=-n_\alpha m_{\alpha\alpha^{\prime}}\nu_{\alpha\alpha^{\prime}}(v_\alpha-v_\alpha^{\prime}),
    \label{eq:3.6}
\end{equation}
where the collisional frequency between various species, $\nu_{\alpha\alpha^{\prime}}$ was defined earlier by equation (\ref{eq:2.2}).

Now let us discuss the momentum transfer for each species. In the case of neutrals, the transfer of momentum will have two terms corresponding to the collision of neutrals with electrons and with ions. Accordingly we have 
\begin{equation}
    P_{ne}=-\frac83\sqrt{2}n_en_nm_e\left(\frac{k_BT}{\pi m_e}\right)^{1/2}\sigma_{ne}(\varv_n-\varv_e),
    \label{eq3.6.1}
    \end{equation}
    \begin{equation}
    P_{ni}=-\frac{8}{3}n_nn_i m_i\left(\frac{k_BT}{\pi m_i}\right)^{1/2}\sigma_{ni}(\varv_n-\varv_i).
    \label{eq:3.6.2}
    \end{equation}
For ions the relevant momentum transfer terms become
    \begin{equation}
        P_{in}=-\frac{8}{3}n_nn_im_i\left(\frac{k_BT}{\pi m_i}\right)^{1/2}\sigma_{in}(\varv_i-\varv_n),
        \label{eq:3.7.1}
    \end{equation}
    and
    \begin{equation}
        P_{ie}=-\frac83\sqrt{2}n_in_em_e\left(\frac{k_BT}{\pi m_e}\right)^{1/2}\sigma_{ie}(\varv_i-\varv_e).
        \label{eq:3.7.2}
    \end{equation}
Finally, for electrons these expressions become
    \begin{equation}
        P_{en}=-\frac83\sqrt{2}n_en_nm_e\left(\frac{k_BT}{\pi m_e}\right)^{1/2}\sigma_{en}(\varv_e-\varv_n),
        \label{eq:3.8.1}
    \end{equation}
    and
    \begin{equation}
     P_{ei}=-\frac83\sqrt{2}n_en_im_e\left(\frac{k_BT}{\pi m_e}\right)^{1/2}\sigma_{ei}(\varv_e-\varv_i).
        \label{eq:3.8.2}
    \end{equation}
We should point out that since we are dealing with elastic collisions, the momentum conservation requires that $P_{\alpha\alpha^{\prime}}+P_{\alpha^{\prime}\alpha}=0$.

Now let us estimate the magnitude of the two terms for each species. Assuming that velocity perturbations are of the same order we have that 
\[
    \left|\frac{P_{ne}}{P_{ni}}\right|={\cal O}\left(\mu^{-1/2}\frac{\sigma_{ne}}{\sigma_{ni}}\right), \quad \left|\frac{P_{in}}{P_{ie}}\right|={\cal O}\left(\frac{n_n}{n_e}\mu^{1/2}\frac{\sigma_{in}}{\sigma_{ie}}\right),
    \]
    \begin{equation}
 \left|\frac{P_{en}}{P_{ei}}\right|={\cal O} \left(\frac{n_n}{n_i}\frac{\sigma_{en}}{\sigma_{ei}}\right).
        \label{eq:3.9}
    \end{equation}
Taking into account that we are dealing with a weakly ionised quasi-neutral hydrogen plasma ($n_i=n_e\ll n_n$) and the typical values of collisional cross sections available in the literature, we can show that $P_{ne}\ll P_{ni}$, $P_{in}\gg P_{ie}$ and $P_{en}\gg P_{ei}$. These orderings between momentum transfer rates will help us simplify the momentum equations for the three species.

The plasma is in hydrostatic equilibrium so that for each species we have
\[
\frac{dp_{0\alpha}}{dz}=-g\rho_{0\alpha}.
\]
In addition, the equilibrium state also satisfies the equations of state, i.e.
\[
p_{0j}=\frac{k_B\rho_{0\alpha}T_{0}}{{\tilde m_{\alpha}}}, 
\]
where $T_{0}$ is the equilibrium values of the  temperature, and ${\tilde m}_{\alpha}$ denotes the mean particle mass for the species $\alpha$. 

As a result, the three simplified momentum equations read
\[
\rho_{0n}\frac{\partial v_n}{\partial t}+\frac{\partial p_n}{\partial z}+\rho_n g=-n_nm_i\nu_{nc}(v_n-v_c),
\]
\[
\rho_{0e}\frac{\partial v_e}{\partial t}+\frac{\partial p_e}{\partial z}+\rho_eg=-n_em_e\nu_{en}(v_e-v_n),
\]
\begin{equation}
    \rho_{0i}\frac{\partial v_i}{\partial t}+\frac{\partial p_i}{\partial z}+\rho_ig=-n_im_i\nu_{in}(v_i-v_n),
    \label{eq:3.10}
\end{equation}
where we used the fact that in a hydrogen plasma $m_i\approx m_n$. Clearly, the evolution of charged particles will be driven by neutrals, which is natural in such weakly ionised and highly collisional plasmas. 
The above equations, together with the mass and energy conservation equations (\ref{eq3.1}, \ref{eq3.5}) will be used to determine the evolutionary equations of waves associated with various species and the property of these waves in the presence of gravitational stratification.

With the simplifications listed above, we can reduce the system of equations to 


\begin{equation}
    \frac{\partial^2 \varv_{e}}{\partial t^2} - c_{S}^2  \frac{\partial^2 \varv_{e}}{\partial z^2} +\gamma g \frac{\partial \varv_{e}}{\partial z}+ \nu_{en} \left(\frac{\partial \varv_{e}}{\partial t}-\frac{\partial \varv_{n}}{\partial t}\right)=0,
    \label{eq20}
\end{equation}
\begin{equation}
    \frac{\partial^2 \varv_{i}}{\partial t^2} - c_{S}^2  \frac{\partial^2 \varv_{i}}{\partial z^2} +\gamma g  \frac{\partial \varv_{i}}{\partial z}+ \nu_{in} \left(\frac{\partial \varv_{i}}{\partial t}-\frac{\partial \varv_{n}}{\partial t}\right)=0,
    \label{eq21}
\end{equation}
\begin{equation}
    \frac{\partial^2 \varv_{n}}{\partial t^2} - c_{S}^2  \frac{\partial^2 \varv_{n}}{\partial z^2} +\gamma g  \frac{\partial \varv_{n}}{\partial z}+ \nu_{ni} \left(\frac{\partial \varv_{n}}{\partial t}-\frac{\partial \varv_{i}}{\partial t}\right)=0.
    \label{eq22}
\end{equation}

Inspired from the results by Alharbi et al. (2021), equations (\ref{eq20})-- (\ref{eq22}) can be written in the standard form of telegrapher's equations for the three species, by introducing new functions of the form
\begin{equation*}
v_{\alpha}(z,t)=\left(\gamma p_{0\alpha}\right)^{-1/2} Q_{\alpha}(z,t).
\end{equation*}
After some straightforward calculations, the three momentum equations can be written as
\begin{equation}
    \frac{\partial^2 Q_e}{\partial t^2}- c_{S}^2  \frac{\partial^2 Q_e}{\partial z^2}+\omega_{e}^2 Q_e =-\nu_{en}\left(\frac{\partial Q_{e}}{\partial t}-\chi^{1/2} \frac{\partial Q_{n}}{\partial t}\right),
    \label{eq23}
\end{equation}
\begin{equation}
    \frac{\partial^2 Q_i}{\partial t^2}- c_{S}^2  \frac{\partial^2 Q_i}{\partial z^2}+\omega_{i}^2 Q_i =-\nu_{in}\left(\frac{\partial Q_{i}}{\partial t}- \chi^{1/2} \frac{\partial Q_{n}}{\partial t}\right),
     \label{eq24}
\end{equation}
\begin{equation}
    \frac{\partial^2 Q_n}{\partial t^2}- c_{S}^2  \frac{\partial^2 Q_n}{\partial z^2}+\omega_{n}^2 Q_n =0,
     \label{eq25}
\end{equation}
where,
\begin{equation*}
    \omega_{\alpha}^2=-\frac{3}{4}\frac{\gamma^2 g^2}{c_{S}^{2}}-\gamma g\frac{d}{dz}\left(\ln \rho_{0\alpha}\right)-\frac{\gamma g}{2} \frac{d}{dz}\left(\ln c_{S}^{2}\right),
\end{equation*}
are the acoustic cut-off frequencies for each species, and $\chi$ is the ionisation fraction defined as $\chi=\rho_{0i}/\rho_{0n}=\mu\rho_{0e}/\rho_{0n}\ll 1$.

It is straightforward to show that the above equations can be written as a form of Klein-Gordon equation by introducing a new function for each species of the form
\[
Q_{\alpha}(z,t)=q_{\alpha}(z,t)\exp(\lambda_{\alpha} t), 
\]
where the values of the parameters $\lambda_{\alpha}$ are chosen such that the first-order derivatives with respect to $t$ vanish. Indeed, by choosing \[
    \lambda_{e}=-\frac{\nu_{en}}{2}, \quad \lambda_{i}=-\frac{\nu_{in}}{2}, \quad 
    \lambda_{n}=-\frac{\nu_{ni}}{2},
\]
the system of equations describing the spatial and temporal evolution of waves for the three species becomes 
\begin{equation}
    \frac{\partial^2 q_e}{\partial t^2}- c_{S}^2  \frac{\partial^2 q_e}{\partial z^2}+{\tilde \Omega}_{e}^2 q_e =\nu_{en}\chi^{1/2} e^{ \left(\nu_{en}-\nu_{ni}  \right) t/2 } \left(\frac{\partial q_{n}}{\partial t}-\frac{\nu_{ni}}{2} q_n \right),
     \label{eq26}
\end{equation}
\begin{equation}
    \frac{\partial^2 q_i}{\partial t^2}- c_{S}^2 \frac{\partial^2 q_i}{\partial z^2}+{\tilde \Omega}_{i}^2 q_i =\nu_{in}\chi^{1/2} e^{ \left(\nu_{in}-\nu_{ni}  \right) t/2 } \left(\frac{\partial q_{n}}{\partial t}-\frac{\nu_{ni}}{2} q_n \right),
     \label{eq27}
\end{equation}
\begin{equation}
    \frac{\partial^2 q_n}{\partial t^2}- c_{S}^2\frac{\partial^2 q_n}{\partial z^2}+{\tilde \Omega}_{n}^2 q_n=0,
     \label{eq28}
\end{equation}
where
\begin{equation}
    {\tilde \Omega}_{e}^2=\omega_e^2-\frac{\nu_{en}^2}{4}, \quad
    {\tilde \Omega}_{i}^2=\omega_{i}^{2}-\frac{\nu_{in}^2}{4}, \quad {\tilde \Omega}_{n}^2=\omega_{n}^{2}-\frac{\nu_{ni}^2}{4},
    \label{cutoffs}
\end{equation}
are the squares of the collision-modified cut-off frequencies for electron, ion, and neutral sound waves, respectively.

\begin{figure}
   	\includegraphics[width=\columnwidth]{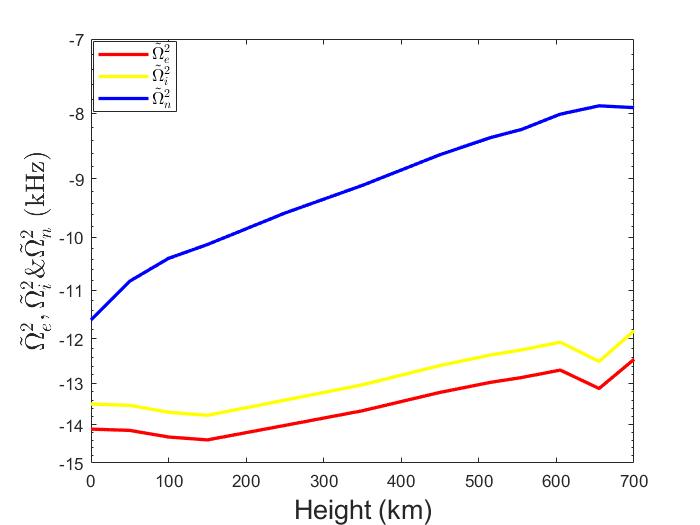}
    \caption{The variation with height of the square of the collision-modified cut-off frequencies for electron, ion, and neutral acoustic-gravity waves based on equations. (\ref{cutoffs}) and assuming a VAL IIIC atmospheric model. The values are given on logarithmic scale.}
         \label{Fig3}
   \end{figure}

In order to estimate the magnitude and importance of these modified cut-off frequencies, we plot their variation with height, where the values of various physical parameters (e.g. densities, temperatures, etc.) are taken from the VAL IIIC atmospheric model (see Figure \ref{Fig3}), with the blue, yellow and red line corresponding to the variation for neutrals, ions and electrons, respectively. It is clear that in the region of the interest (but true for the whole weakly ionised solar atmosphere) these quantities are negative, meaning that collisional effects are more important than stratification effects and collisions between particles will prevent any growth of waves' amplitudes. This result implies that gravitational effects can be confidently neglected when discussing the propagation of waves in a multi-fluid framework, so the dynamics of waves can be described within the framework of a homogeneous (non-stratified )plasma.
As a result, the ${\tilde \Omega}_\alpha^2$ coefficients of the last terms on the left-hand side of equations (\ref{eq26})--(\ref{eq28}) will be negative, therefore, waves will propagate with no cut-off frequency.
 

 
  
We can carry out a normal mode analysis for the homogeneous part of equations (\ref{eq26}-\ref{eq28}), by assuming perturbations proportional to $e^{i(kz-\omega t)}$, where $k$ is the wavenumber of waves and $\omega$ is the frequency. Now the governing equations for the three species reduce to the dispersion relations
 \begin{equation}
  \omega^2=k^2 c_{S}^2-{\tilde \Omega}_{\alpha}^2.  \label{eq:29}
 \end{equation}
 Clearly, waves will propagate (i.e. $\omega^2 > 0$) provided the wavenumbers attached to these waves satisfy the conditions
\begin{equation}
     k_c^e>\frac{{\tilde \Omega}_{e}}{c_{S}} \approx \frac{\nu_{en}}{2 c_{S}},\; k_c^i>\frac{{\tilde \Omega}_{i}}{c_{S}} \approx \frac{\nu_{in}}{2c_{S}},\;
      k_c^n>\frac{{\tilde \Omega}_{n}}{c_{S}} \approx \frac{\nu_{ni}}{2 c_{S}},
      \label{eq:29.1}
 \end{equation}
where the quantities $k_c^{\alpha}$ constitute the values of the wavenumber cut-off corresponding to the three species.


With the gravitational effects safely neglected we can treat the problem of dynamics in a homogeneous plasma, allowing us to discuss a more complex model. Without loss of generality, we can assume a two dimensional propagation in the $xz$ plane. In this case the governing linearised equations (\ref{eq:3.1.1})--(\ref{eq:3.1.3}) for the three species become
\begin{equation} \label{eq33}
     \frac{\partial \rho_{\alpha}}{\partial t}+\rho_{0\alpha} \left(\frac{\partial v_{x \alpha}}{\partial x}+ \frac{\partial v_{z \alpha}}{\partial z}\right)=0,
 \end{equation}

\begin{equation}\label{eq34}
    \rho_{0\alpha} \frac{\partial v_{x \alpha}}{\partial t}+ \frac{\partial p_{\alpha}}{\partial x}=\sum_{\alpha^{\prime}\neq \alpha}P_{\alpha\alpha^{\prime}},
\end{equation}

\begin{equation}\label{eq35}
    \rho_{0\alpha} \frac{\partial v_{z \alpha}}{\partial t}+ \frac{\partial p_{\alpha}}{\partial z}=\sum_{\alpha^{\prime}\neq \alpha}P_{\alpha\alpha^{\prime}},
\end{equation}

\begin{equation}  \label{eq36}
    \frac{\partial p_{\alpha}}{\partial t}=c_{S}^2 \frac{\partial \rho_{\alpha}}{\partial t}.
\end{equation}

After long but straightforward calculations, the system of governing equations can be reduced to a system of partial differential equations for each species
\begin{equation}\label{eq37}
 \left(c_{S}^2 \frac{\partial^2}{\partial x^2}  -\frac{\partial^2}{\partial t^2} - \nu_{\alpha\alpha^{\prime} } \frac{\partial}{\partial t}\right) v_{x \alpha} + c_{S}^2 \frac{\partial^2 v_{z \alpha}}{\partial x\partial z}+ \nu_{\alpha \alpha^{\prime}} \frac{\partial v_{x \alpha^{\prime}}}{\partial t}=0,
\end{equation}
\begin{equation}
    \left(c_{S}^2 \frac{\partial^2}{\partial z^2}-\frac{\partial^2}{\partial t^2}- \nu_{\alpha \alpha^{\prime}} \frac{\partial}{\partial t} \right) v_{z \alpha}+  c_{S}^2  \frac{\partial^2 v_{x \alpha}}{\partial x\partial z}+\nu_{\alpha \alpha^{\prime}} \frac{\partial v_{z \alpha^{\prime}}}{\partial t}=0.
    \label{eq38}
\end{equation}





Next, in the above two equations we apply a Fourier analysis, writing all velocity components proportional to $\exp(i k_x+i k_z-i\omega t)$. Using the compatibility condition of the system of equations written for the amplitude of velocities, we obtain the dispersion relation of the three acoustic modes in the form of a sixth order polynomial of the form 
\[
    \left(-\omega^2+k^2 c_S^2-i\omega \nu_{en} \right)  \Big[\omega^4 +i \omega^3 (\nu_{in}+\nu_{ni})  -\omega^2 (\nu_{in} \nu_{ni} -\nu_{ni}^{2}+2 k^2 c_S^2)
\]
\begin{equation}
   -i \omega k^2 c_S^2(\nu_{in}+\nu_{ni})+ k^4 c_S^4 \Big]=0.
   \label{eq:42}
\end{equation}

\begin{figure}
   	\includegraphics[width=\columnwidth]{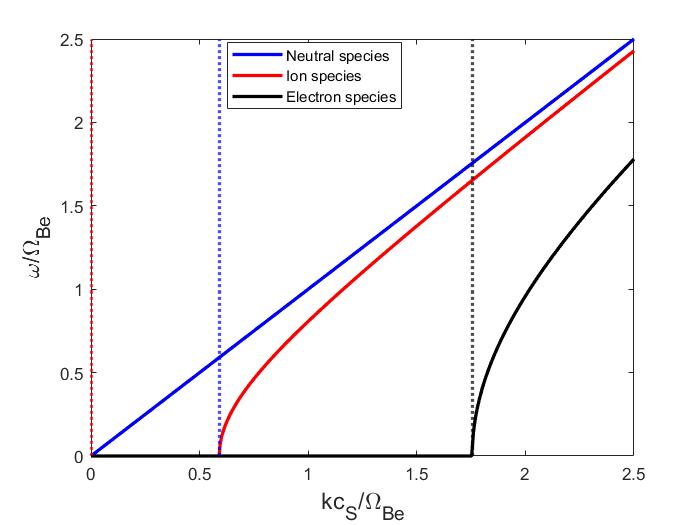}
    \caption{The real part of frequencies of modes (in units of the electron gyro-frequency, $\Omega_{Be}$) given by equation (\ref{eq:42}) in terms of the dimensionless variable $kc_S/\Omega_{Be}$ for the sound waves associated to the three species. Here the variation of the frequency for neutral, ion and electron sound waves are given by blue, red and black solid lines, respectively).}
         \label{Fig4}
   \end{figure}

\begin{figure}
   	\includegraphics[width=\columnwidth]{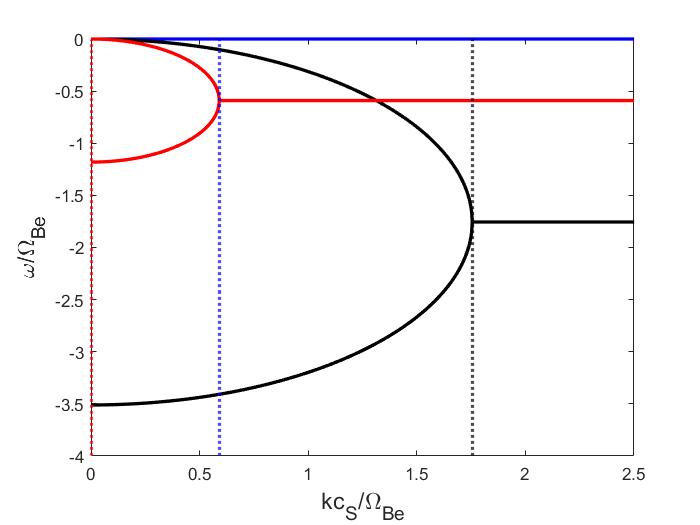}
    \caption{The same as in Figure \ref{Fig4}, but here we plot the variation of the imaginary part of the frequencies, as given by equation (\ref{eq:42}).}
         \label{Fig5}
   \end{figure}
The variation of the real and imaginary parts of the dimensionless frequency (in units of the electron gyro-frequency) with respect to the dimensionless frequency $kc_S/\Omega_{Be}$ are shown in Figures \ref{Fig4} and \ref{Fig5} for the acoustic modes associated to the neutral, ion and electron species shown by blue, red and black solid lines, respectively. To produce these figures the characteristic values of physical parameters were chosen to be $c_S=10$ km s$^{-1}$, $T=4465$ K, $\nu_{in}=1.8\times 10^8$ Hz, $\nu_{ni}=2\times 10^4$ Hz, $\nu_{en}=5.5\times10^8$ Hz and $\Omega_{Be}=1.5\times 10^8$ Hz. The cut-off values given by equation (\ref{eq:29.1}) are represented as vertical dotted lines. In practice these plots display the variation of the dimensionless frequencies with the wavenumber (or its inverse, the wavelength). It is clear that in this region the sound waves associated to neutrals propagate practically with no cut-off value and their frequency increases with decreasing the wavelength. Their dispersion curve follows the $\omega=kc_S$ line and these waves propagate practically with no damping (see Figure \ref{Fig5}). 

Given the high collisional frequency between ions and neutrals, the frequency of ion sound waves quickly becomes equal to the frequency of neutral sound waves. For large values of the wavelength (small values of $kc_S/\Omega_{Be}$ sound waves associated with ions propagate only if their wavenumber is larger than a critical value that can be defined as $k\geq \nu_{in}/2c_S\approx 9.2$ km$^{-1}$. Below the critical wavenumber ion sound waves are simple entropy modes (Goedbloed and Poedts 2004, Murawski et al. 2011, Soler et al. 2013b), i.e. non-propagating. The frequency of entropy modes are purely imaginary and involve perturbations in the plasma density (or internal energy), but not in pressure. The transition of these waves from entropy, non-propagating waves, to acoustic propagating waves occurs at the critical wavenumber. For propagating ion sound waves their frequency increases with decreasing the wavelength and their damping rate is independent on the wavelength of waves.    
Finally, the sound waves associated with electrons have a much higher cut-off value and will propagate when $k\geq \nu_{en}/2c_S\approx 27.3$ km$^{-1}$. Although the electron-related sound waves show initially a distinct behaviour compared to the other two modes (due to the increased mobility of electrons), for decreasing wavelengths, the frequency of these waves will tend to the frequency of neutral sound waves. The imaginary part of the electron sound waves is shown in Figure \ref{Fig5} (black lines). Similar to the ion sound waves, as long as the wavenumber is smaller than the critical value, these waves are non-propagating, entropy waves. For wavenumbers larger than the critical value, waves have a wavelength-independent strong damping. Comparing the magnitude of the real and imaginary parts of the frequency it is obvious that while initially these waves have a strong damping, for smaller wavelengths the waves become weakly damped.

The important result of these figures is that the collisional frequencies play an essential role in the propagation of waves. Waves will be damped with rate that depend only on collisional frequency. In addition, the critical wavenumbers of waves are determined by the strength of collisions between species.  

\subsection{Waves in Region II}

The variation of the collisional and gyro-frequencies with height shown in Figure \ref{frequencies} reveals that at heights roughly corresponding to the base of the chromosphere, the collisional frequency of electrons falls below the electron gyro-frequency, therefore, electrons become magnetised. At the same time the collisional frequency of ions is still above the ion gyro-frequency, meaning that ions are still not affected by the magnetic field and their dynamics of predominantly driven by the collisions with neutrals. As a result, the dynamics of electrons and ions becomes different, however they form a single fluid since the collisional frequency of electrons with ions is still very strong (see Figure \ref{Fig2}). In frequency domain Region II corresponds to $\Omega_{Bi}< \omega<\Omega_{Be}$, where $\omega_{Be,i}$ were defined earlier. The strong collisional coupling between neutrals and ions prevents ions to magnetise. Electrons moving along the magnetic field, drift due to ${\bf E}\times {\bf B}$ in the transversal direction, while ions will continue to move together with neutrals. Here ${\bf E}$ is the electric field generated as a result of the different motion of electrons and ions. 

The disassociation of charged particles results in generation of an electric current ${\bf J}=en_e({\bf v}_i-{\bf v}_e)$. This current will generate an electric force ($en_e{\bf E})$ aimed to oppose the movement of ions. 

We assume a uni-directional homogeneous magnetic field of the form ${\bf B}_0=B_0\hat{\bf z}$. Based on the results we obtained earlier, gravitational effects will be neglected.  


Since we are dealing with a two-fluid plasma, the momentum equation for charged particles can be obtained by combining the linearised momentum equations of electrons and ions
\begin{equation}
    \rho_{0e}\frac{\partial {\bf v}_e}{\partial t}+\nabla p_e=-en_e({\bf E}+{\bf v}_e\times {\bf B}_0)-\rho_e\nu_{en}({\bf v}_e-{\bf v}_n),
    \label{eq:4.1}
\end{equation}
\begin{equation}
    \rho_{0i}\frac{\partial {\bf v}_i}{\partial t}+\nabla p_i=+en_i{\bf E}-\rho_e\nu_{in}({\bf v}_i-{\bf v}_n).
    \label{eq:4.2}
\end{equation}
As a result, the momentum equation for charges becomes
\begin{equation}
    \rho_{0c}\frac{\partial {\bf v}_c}{\partial t}+\nabla p_c=-en_e{\bf v}_e\times {\bf B}_0-\rho_{0c}\nu_{cn}({\bf v_c}-{\bf v_n}),
    \label{eq:7.2.1}
\end{equation}
where
\[ 
\rho_{0c}=\rho_{0e}+\rho_{0i}, \quad {\bf v}_c=\frac{\rho_{0i}{\bf v}_i+\rho_{0e}{\bf v}_e}{\rho_{0c}},
\]
\[
\nu_{cn}=\frac{\rho_{0i}\nu_{in}+\rho_{0e}\nu_{en}}{\rho_{0c}},
\]
are the total density of charged species, the velocity of the center of the mass, and the collisional frequency of charged fluid weighed by the density of the two species. Finally, the momentum conservation of neutrals can be simply written as
\begin{equation}
  \rho_{0n}\frac{\partial {\bf v}_n}{\partial t}+\nabla p_n=-\rho_{0n}\nu_{nc}({\bf v}_n-{\bf v}_c).
    \label{eq:7.2.2}
\end{equation}

Despite the two-fluid approximation, only electrons are magnetized and, through collisions, electrons are perturbing the ion population. We should also keep in mind that neutrals and ions are still coupled, which results in a strong drag force resulting in the bulk of ion velocity, ${\bf v}_i$ closely matching the velocity of the neutral gas. With the charged particles having different motion (electrons gyrating around the equilibrium magnetic field and ions having a rectilinear motion advected by neutrals) the quasi-neutrality of the plasma is going to be perturbed, however the electric field generated will always aim to restore the quasi-neutrality of the system. Ions are still more strongly coupled to neutrals than to electrons since
\[
\frac{\nu_{in}}{\nu_{ie}}\approx \frac{n_n}{n_i}\frac{\sigma_{in}}{\sigma_{ie}}\mu^{3/2}\gg 1.
\]
Therefore, the current can be written as $
{\bf J}\approx en_e({\bf v}_n-{\bf v}_e)$
or
\begin{equation}
{\bf v}_e={\bf v}_n-\frac{{\bf J}}{en_e}.
\label{eq:7.2.3}
\end{equation}
The role of the current generated in this layer of the solar atmosphere thanks to the disassociation of charged particles was discussed in details in an earlier study by Krasnoselskikh et al. (2010), who showed that these currents are formed around the layer where electron gyrofrequency and collisional frequency are comparable. Currents can also effectively dissipate via ohmic dissipation resulting in typical temperature increases with altitude as large as $0.1-0.3$ eV km$^{-1}$. The heat generated as a result of dissipation can create additional ionisation of particles, leading to the modification in the thermal equilibrium of the plasma. In addition, the currents can generate local magnetic fields (acting as dynamos) comparable with the background magnetic field. The idea of current dissipation in the partially ionised solar atmosphere is relatively new aspect of solar atmospheric research, and remains to be seen how much the dissipation of these currents contribute to the global process of chromospheric plasma heating.

Taking into account the Amp\'ere's law, we can write the momentum equation for the charged species as
\begin{equation}
 \rho_{0c}\frac{\partial {\bf v}_c}{\partial t}+\nabla p_c=-en_e{\bf v}_n\times {\bf B}_0+\frac{1}{\mu_0}(\nabla\times {\bf b})\times {\bf B}_0-\rho_{0c}\nu_{cn}({\bf v_c}-{\bf v_n}).
    \label{eq:7.2.4}
\end{equation}  
Since the magnetic field is present in the above equation, we will also need to derive the induction equation from the momentum equation for electrons, assuming that the electron inertia and electron pressure can be neglected. As a result we have
\begin{equation}
    en_e({\bf E}+{\bf v}_e\times {\bf B})=(\nu_{ei}+\nu_{en})\frac{m_e}{e}{\bf J},
    \label{eq:7.2.5}
\end{equation}
where we took into account equation (\ref{eq:7.2.3}). Using the same equation on the left hand side of the above relation, we obtain 
\begin{equation}
    {\bf E}+{\bf v}_n\times {\bf B}-\frac{1}{en_e}{\bf j}\times {\bf B}=\nu_e\frac{m_e}{e^2n_e}{\bf J},
    \label{eq:7.2.6}
\end{equation}
where $\nu_e=\nu_{en}+\nu_{ei}$ is the total collisional frequency of electrons. After taking the curl of the above equation and using Faraday's law of induction ($\nabla\times {\bf E}=-\partial {\bf B}/\partial t$), we obtain
\begin{equation}
    \frac{\partial {\bf B}}{\partial t}=\nabla\times ({\bf v}_n\times {\bf B})-\frac{1}{en_e}\nabla\times ({\bf j}\times {\bf B})-\frac{m_e\nu_e}{e^2n_e}\nabla\times {\bf J},
    \label{eq:7.2.7}
\end{equation}
which means that when ions and neutrals are strongly coupled, the neutrals get magnetized. The last two terms of the above equation are the Hall term that renders waves to be dispersive on scales of the order of the ion inertial length and the resistive term that contains, as coefficient, the resistivity due to electrons moving along the magnetic field (and describes the dissipation of field-aligned currents). The currents in this model can dissipate through the collisions of electrons not only with ions, but also with neutrals.

The governing equations in Region II will be the mass conservation equations for charged particles and neutrals given by equation \ref{eq:3.1.1}, the momentum equations for ions and neutrals (\ref{eq:7.2.1} and \ref{eq:7.2.2}), the induction equation (\ref{eq:7.2.5}) and the energy equations for the two species (similar to the linearised equation \ref{eq:3.1.3}), considering, again, an adiabatic process.  
Similar to the waves studied in Region I, we assume a two-dimensional dynamics and write ${\bf k}=(k_x,0,k_z)$ or ${\bf k}=(k\sin\theta,0,k\cos\theta)$, where $\theta$ is the angle between the direction of propagation and the $z$ axis. We also assume waves oscillating in time, so we consider all perturbations proportional to $e^{-i\omega t}$, where $\omega$ is the complex frequency.

The combination of all these equations results in the system of equations written for the components of velocity, as

\begin{figure*}
  \subfloat{
	\begin{minipage}[c][1\width]{
	   0.3\textwidth}
	   \centering
	   \includegraphics[width=1\textwidth]{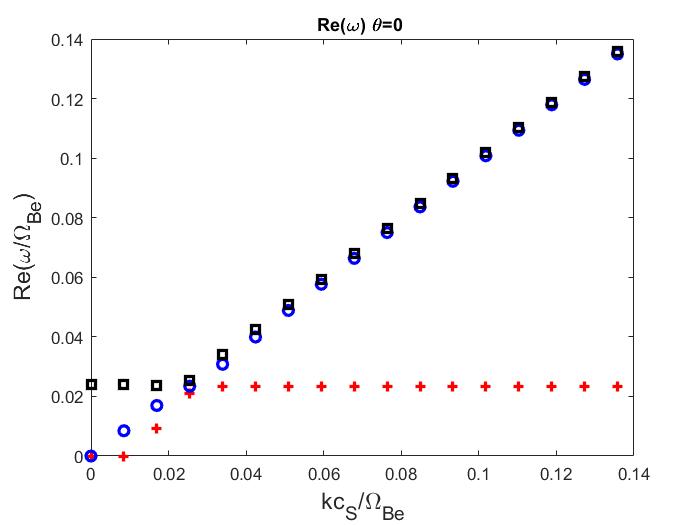}
	\end{minipage}}
 \hfill 	
  \subfloat{
	\begin{minipage}[c][1\width]{
	   0.3\textwidth}
	   \centering
	   \includegraphics[width=1.1\textwidth]{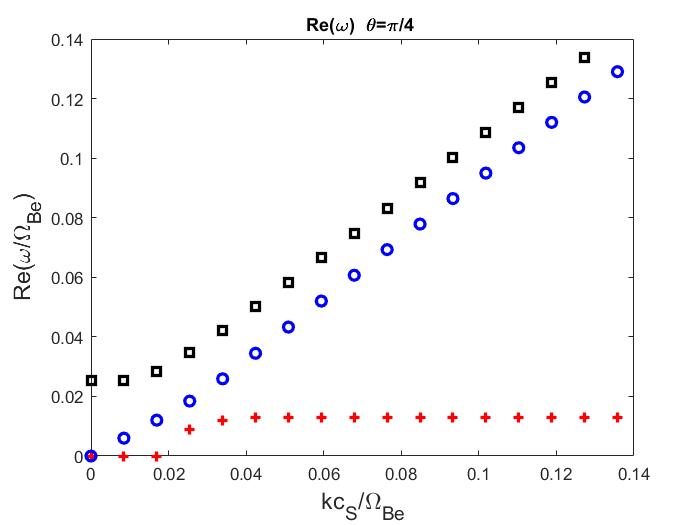}
	\end{minipage}}
 \hfill	
  \subfloat{
	\begin{minipage}[c][1\width]{
	   0.3\textwidth}
	   \centering
	   \includegraphics[width=1.2\textwidth]{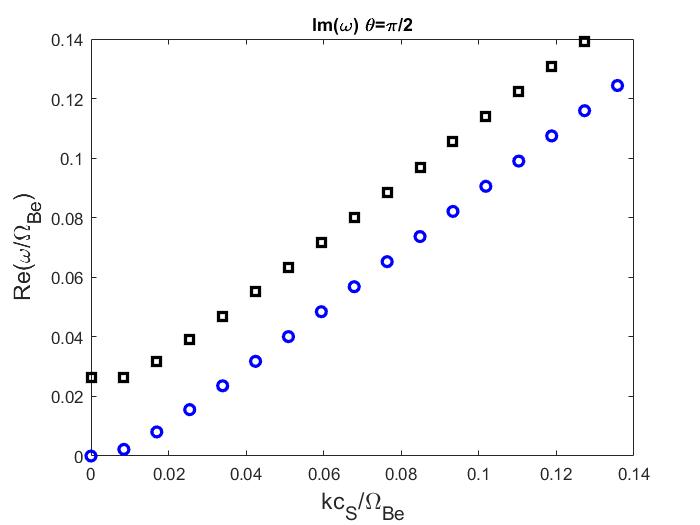}
	\end{minipage}}
\caption{Real part of the frequency (in units of the electron gyro-frequency, $\Omega_{Be}$) in terms of the dimensionless frequency $kc_S/\Omega_{Be}$, as solution of the system of equations (\ref{eq:7.3})-(\ref{eq:7.6}). The three panel correspond to a parallel propagation (left panel), waves propagating at a $\pi/4$ angle with respect to the magnetic field (central panel) and perpendicular to the field (right panel). The coloured curves correspond to charged slow waves (red), neutral slow wave (blue) and fast waves (black).}
\label{realomega}
\end{figure*}

\begin{figure*}
  \subfloat{
	\begin{minipage}[c][1\width]{
	   0.3\textwidth}
	   \centering
	   \includegraphics[width=1\textwidth]{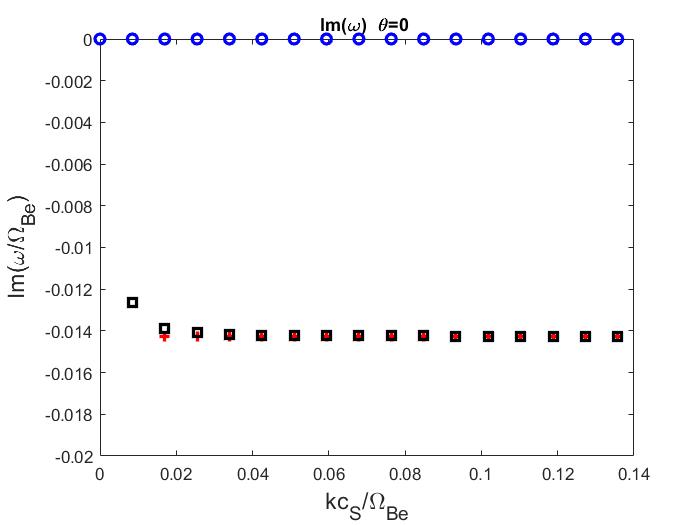}
	\end{minipage}}
 \hfill 	
  \subfloat{
	\begin{minipage}[c][1\width]{
	   0.3\textwidth}
	   \centering
	   \includegraphics[width=1.1\textwidth]{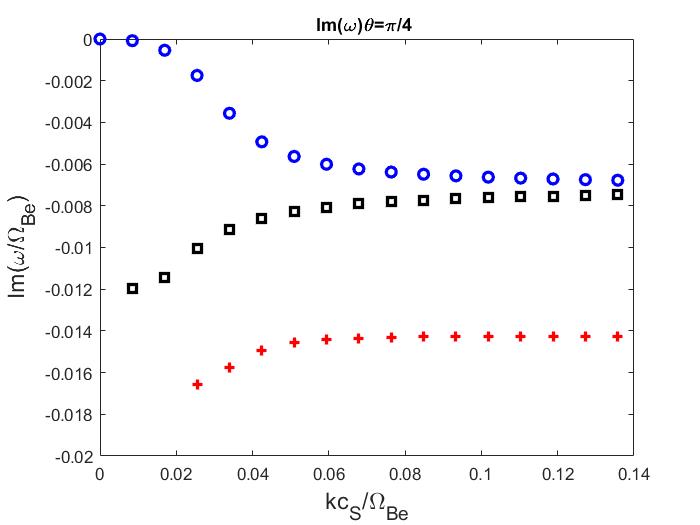}
	\end{minipage}}
 \hfill	
  \subfloat{
	\begin{minipage}[c][1\width]{
	   0.3\textwidth}
	   \centering
	   \includegraphics[width=1.2\textwidth]{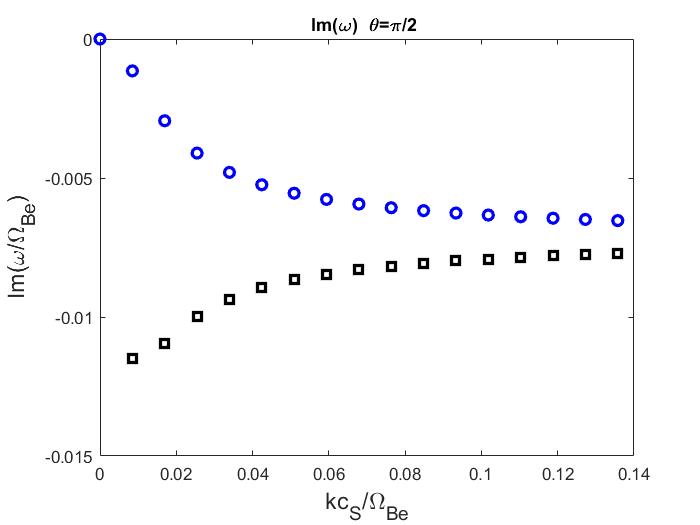}
	\end{minipage}}
\caption{The same as Figure \ref{realomega}, but here we plot the imaginary parts of the frequency. The colours correspond to the type of waves defined in Figure \ref{realomega}.}
\label{imaginaryomega}
\end{figure*}
\begin{equation}
(\omega^2-k_x^{2} c_S^{2}+i \chi \omega \nu_{cn}) v_{nx} -k_x k_z c_S^{2} v_{nz}-i \chi \omega \nu_{cn} v_{cx}=0,
\label{eq:7.3}
\end{equation}
\begin{equation}
(\omega^2-k_z^{2} c_S^{2}+i \chi \omega \nu_{cn}) v_{nz} -k_x k_z c_S^{2} v_{nx}-i \chi \omega \nu_{cn} v_{cz}=0,
\label{eq:7.4}
\end{equation}
\[
(\omega^2-k_x^{2} c_S^{2}+i \omega \nu_{cn}) v_{cx} -k_x k_z c_S^{2} v_{cz}-i\omega\left(\frac{ k_z^2 c_{A}^2 }{i\omega -\eta k^2}+ \nu_{cn}\right) v_{nx}
\]
\begin{equation}
+\frac{i \omega k_xk_z c_{A}^2}{i \omega -\eta k^2} v_{nz}=0,
\label{eq:7.5}
 \end{equation}
  \[
(\omega^2-k_z^{2} c_S^{2}+i \omega \nu_{cn}) v_{cz} -k_x k_z c_c^{2} v_{cx}-i\omega\left(\frac{k_x^2 c_{A}^2 }{i\omega -\eta k^2}+ \nu_{cn}\right) v_{nz}
   \]
   \begin{equation}
       +\frac{i \omega k_xk_z c_{A}^2 }{i \omega -\eta k^2} v_{nx}=0,
     \label{eq:7.6}
 \end{equation}
where the constant $\eta$ is the resistivity coefficient, valid in a partially ionised plasmas, and it is defined as
\[
     \eta=\frac{\nu_e m_e}{n_i\mu_0 e^2},
\]
and $c_A=B_0/\sqrt{\mu_0\rho_{0c}}$ is the Alfv\'en speed of the charged species.
 
The above homogeneous system of equations admits solutions only if the determinant of the matrix constructed by means of the coefficients multiplying the components of velocity for the two fluids is vanishing. The dispersion relation is solved numerically and the real and imaginary solutions are represented in Figures \ref{realomega} and \ref{imaginaryomega}. The real and imaginary parts of the frequency is plotted in units of the electron gyro-frequency ($\Omega_{Be}$) as a function of the dimensionless quantity $kc_S/\Omega_{Be}$ for the characteristic values in Region II: $T=5650$ K, $B=24.05$ G, $n_i=1.06\times 10^{17}$ m$^{-3}$, $v_A=206.83$ km s$^{-1}$, $c_S=11.42$ km s$^{-1}$, $\nu_{in}=1.28\times 10^6$ Hz and $\nu_e=2.57\times 10^7$ Hz. In these figures the possible modes are represented by symbols of different colours, so red crosses describe the solution of the dispersion relation corresponding to slow waves associated to charges, the slow wave associated to neutrals is shown by blue circles, and the fast mode is shown by black squares. The three panels correspond parallel propagation ($\theta=0$), a $\pi/4$ inclined propagation with respect to the ambient magnetic field, and a perpendicular propagation, respectively. 

\begin{figure*}
  \subfloat{
	\begin{minipage}[c][1\width]{
	   0.3\textwidth}
	   \centering
	   \includegraphics[width=1\textwidth]{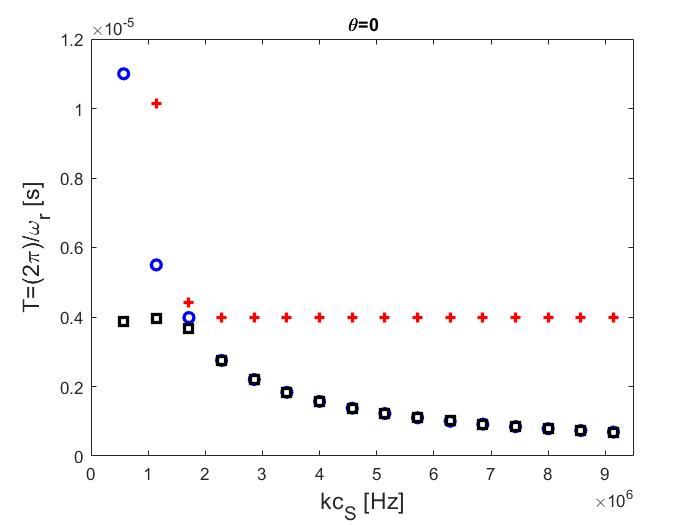}
	\end{minipage}}
 \hfill 	
  \subfloat{
	\begin{minipage}[c][1\width]{
	   0.3\textwidth}
	   \centering
	   \includegraphics[width=1.1\textwidth]{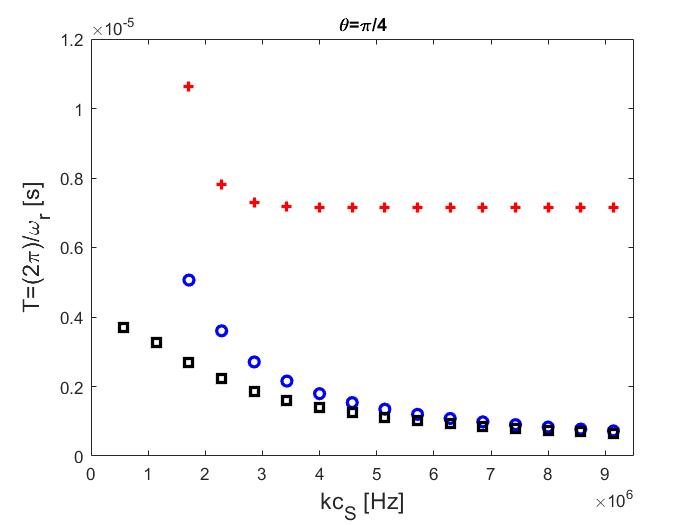}
	\end{minipage}}
 \hfill	
  \subfloat{
	\begin{minipage}[c][1\width]{
	   0.3\textwidth}
	   \centering
	   \includegraphics[width=1.2\textwidth]{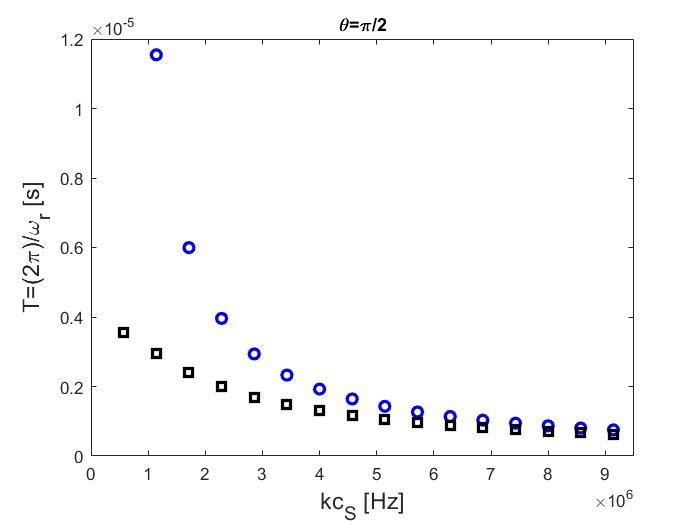}
	\end{minipage}}
\caption{The same waves as in Figure \ref{realomega}, but here we display the period of waves in terms of the frequency of sound waves, $kc_S$. }
\label{period}
\end{figure*}

\begin{figure*}
  \subfloat{
	\begin{minipage}[c][1\width]{
	   0.3\textwidth}
	   \centering
	   \includegraphics[width=1\textwidth]{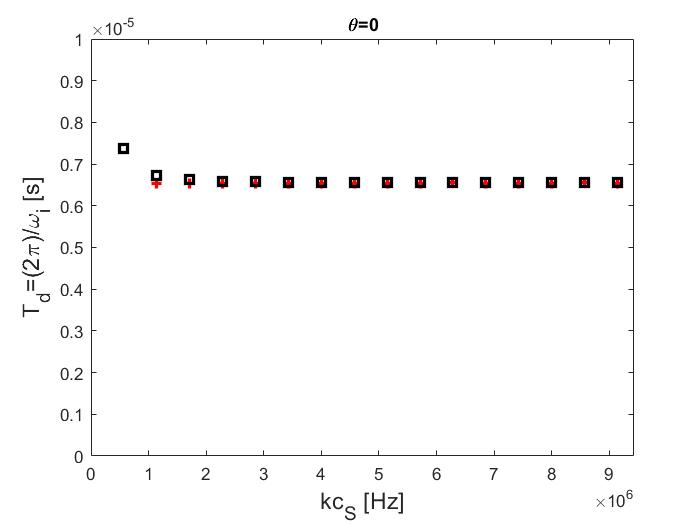}
	\end{minipage}}
 \hfill 	
  \subfloat{
	\begin{minipage}[c][1\width]{
	   0.3\textwidth}
	   \centering
	   \includegraphics[width=1.1\textwidth]{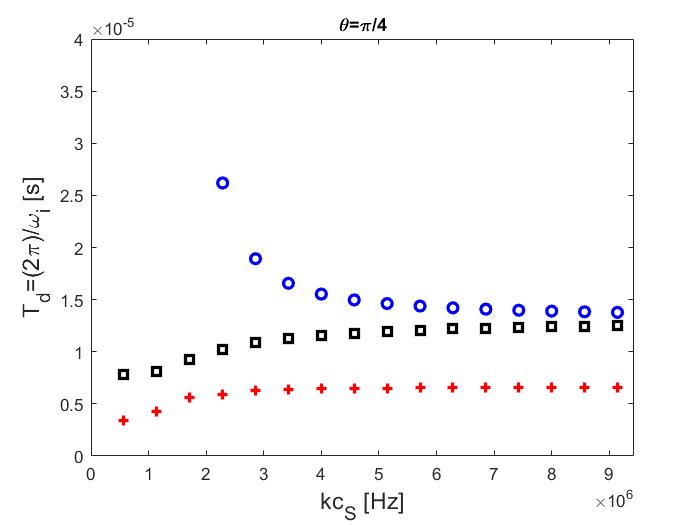}
	\end{minipage}}
 \hfill	
  \subfloat{
	\begin{minipage}[c][1\width]{
	   0.3\textwidth}
	   \centering
	   \includegraphics[width=1.2\textwidth]{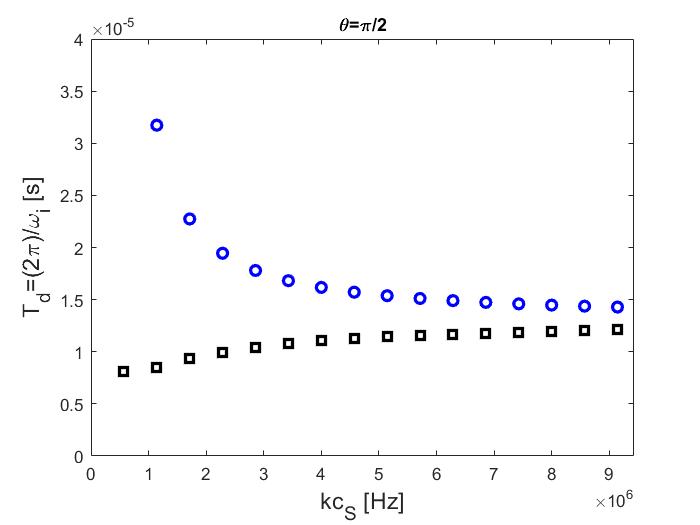}
	\end{minipage}}
\caption{The same waves as in Figure  \ref{imaginaryomega}, but here we show the damping times of waves in terms of the frequency of sound waves, $kc_S$.}
\label{dampingtime}
\end{figure*}

The behaviour of the three waves shows a strong dependence with the propagation direction. In the case of parallel propagation for very small values of wavenumber, $k$, (very large wavelengths) the waves that propagate are the fast waves that have a phase speed determined mainly by the Alfv\'en speed and the neutral slow waves. Due to the strong coupling on neutrals and ions the frequency of these two waves is equal as the wavelength is decreasing. The slow waves associated to charged particles has a cut-off wavenumber and for decreasing wavelength the frequency of these waves becomes independent on their wavelength, i.e. they become non-dispersive. 

The damping rate of waves propagating parallel to the magnetic field is shown in the left-hand panel of Figure \ref{imaginaryomega}. While the slow waves associated to the neutral particles propagate practically with no damping, the damping rates of slow waves of charged species and the fast waves are identical and show very little dependence on the wavelength of waves. 

When the propagation direction is such that the wavevector makes an angle of $\pi/4$ with respect to the ambient magnetic field the slow waves due to the charged species has, again, a cut-off value and the propagation speed increases inversely proportional to the wavelength of waves. Similar to the parallel propagation, at a value of $kc_S/\Omega_{Be}\approx 0.05$, the speed saturates and becomes independent on the wavelength of the waves. Compared to the case of parallel propagation, these waves have a lower frequency. 

Once the waves propagate in a non-parallel direction to the magnetic field, the fast and neutral-associated slow waves decouple. Their frequency is close to each other and they follow the same dependence on the wavelength. As before, frequency of the neutral slow waves increases monotonically along the direction $\omega=kc_S$. The damping rates of these waves are shown in the middle panel of Figure \ref{imaginaryomega}. In the case of charged-related slow waves the damping rate becomes again independent on the wavelength of waves for $kc_S/\Omega_{Be}\approx 0.07$. These waves posses the largest damping rate. The damping rate of fast waves has a similar behaviour as in the case of parallel propagation and, again, it tends to saturate so that it becomes independent on the wavelength of waves. The damping rate of slow waves associated with neutrals has an interest behaviour since for large wavelengths, the damping rate of these waves is zero, i.e. these waves propagate with no damping. The damping rate of these waves increases, and saturates with decreasing the wavelength of waves. The damping rate of fast waves damp quickly for large wavelengths, and their damping rate decreases with the wavelength. For decreasing wavelengths the damping rate of fast waves also saturates and tends to be equal to the damping rates of slow waves associated to neutrals. 

For propagation perpendicular to the background equilibrium magnetic field slow waves associated with charged particles ceases to exist, instead neutral slow waves and fast waves can propagate with frequencies that increase with decreasing wavelength. Similar to the previous cases, and the frequency separation between these two modes increases. The imaginary part of the frequency of these two waves show that with decreasing the wavelength the damping rate of neutral slow wave increases, while for fast waves, it decreases. For smaller wavelengths the damping rate of the two modes becomes almost independent on the wavelength of waves and they tend to converge towards the same value. 

We have repeated these plots for higher heights in the solar chromosphere and the results show that the pattern of the real and imaginary parts of the frequencies shown in Figures (\ref{realomega}) and (\ref{imaginaryomega}) is preserved, the only change is a change of the values. Accordingly, with height, the the real part of frequencies for neutral slow and fast waves are practically unaffected, however they tend to be closer to each other with the increase with height. In contrast the values of the frequencies for slow waves associated to charges decreases. The imaginary part of the frequencies corresponding to all three waves tends to decrease, meaning that with height these waves tend to have weaker damping. 

The period and the damping times of waves corresponding to the field and plasma parameters mentioned earlier are shown in Figures \ref{period} and \ref{dampingtime}. Here these quantities are plotted in terms of the frequency of sound waves, $kc_S$, or, since the sound speed is constant, in terms of the wavenumber, $k$. Apart from the period of slow waves associated with the charged particles, where the period becomes constant for decreasing wavelength, the period of the other two ways decreases with the wavelength. In all cases the periods are of the order of $10^{-6}$, i.e. the collisonal time between ions and neutrals. 

When looking at the damping times shown in Figure \ref{dampingtime}, the curve corresponding to neutral slow waves propagating along the field is not shown because these waves propagate practically undamped. For any propagation angle the damping time of waves becomes gradually independent on the wavelength of waves, when their wavelength decreases.  

\begin{figure*}
   \centering
   \begin{subfigure}{.49\textwidth}
        \centering
        \includegraphics[width=9.cm]{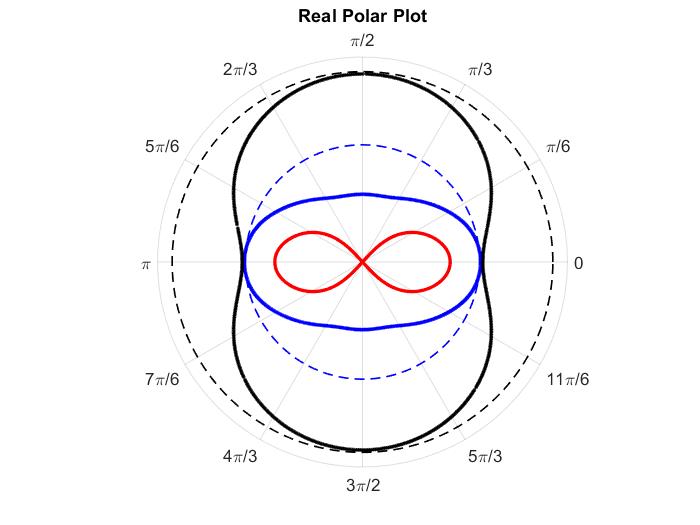}
       \end{subfigure}
   \begin{subfigure}{.49\textwidth}
        \centering
        \includegraphics[width=9.cm]{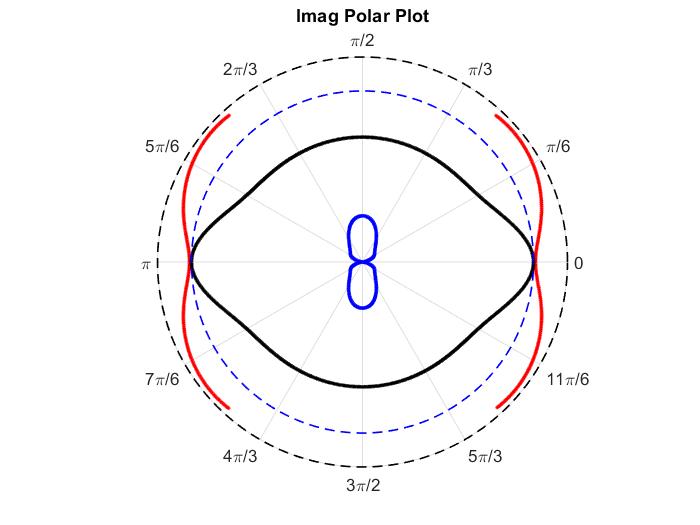}
         \end{subfigure}   
\caption{The polar (Friedrich) diagram of the real (left panel) and imaginary part (right panel) of the dispersion relation. Here the background magnetic field is along the $x$ axis and the direction of waves' propagation is covering a whole $2\pi$ range. The colours are representing the same modes as defined in Figure \ref{realomega}. Note that in the right-hand side panel we plot the absolute value of the damping rate. The dashed lines correspond to the reference values of frequency and damping rate given in the text of the article. These figures were obtained for $kc_S/\Omega_{Be}=0.035$}
   \label{polar}
   \end{figure*}

A full picture of the propagation speed of the waves and their damping rate can be obtained by plotting these values in a Friedrich polar diagram (see Figure \ref{polar}), where the equilibrium magnetic field points in the $x$-direction and the angle of propagation of waves covers a full $2\pi$ range. The diagrams shown in Figure \ref{polar} were obtained by assuming $kc_S/\Omega_{Be}=0.035$ and a plasma-$\beta$ parameter of $3\times 10^{-3}$. In these plots the modes are represented by the same colours as in Figures \ref{realomega} and \ref{imaginaryomega}. The dashed in the polar plots are used for reference values. For the real part of the frequency the blue and black dashed line correspond to 2 and 3.25 MHz, respectively. In the case of the imaginary part of the frequency the dashed blue and black lines correspond to 1.33 and 1.6 MHz. 

From these diagrams it is obvious that fast waves propagate with the highest frequency in the perpendicular direction to the magnetic field, a similar characteristic as fast magnetoacoustic waves in fully ionised plasmas. In the solar atmospheric region we are investigating there are two slow waves associated to the two species, but the slow wave associated with the neutral species propagates always with a higher speed than the one that corresponds to charges. Due to the strong coupling between ions and neutrals, in the case of parallel propagation, the fast waves and slow waves due to neutrals propagate with the same frequency. For parallel propagation the frequency of neutral slow waves is always larger than the phase speed of slow waves associated with the charged particles.

    \begin{figure*}
   \centering
   \begin{subfigure}{.49\textwidth}
        \centering
        \includegraphics[width=9.cm]{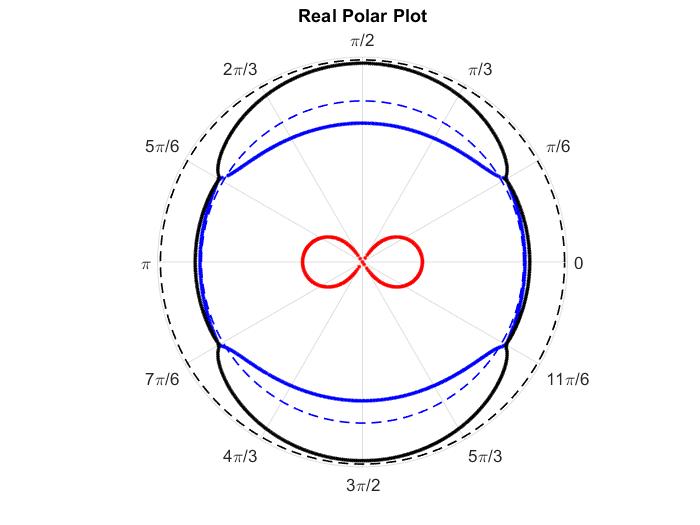}
       \end{subfigure}
   \begin{subfigure}{.49\textwidth}
        \centering
        \includegraphics[width=9.cm]{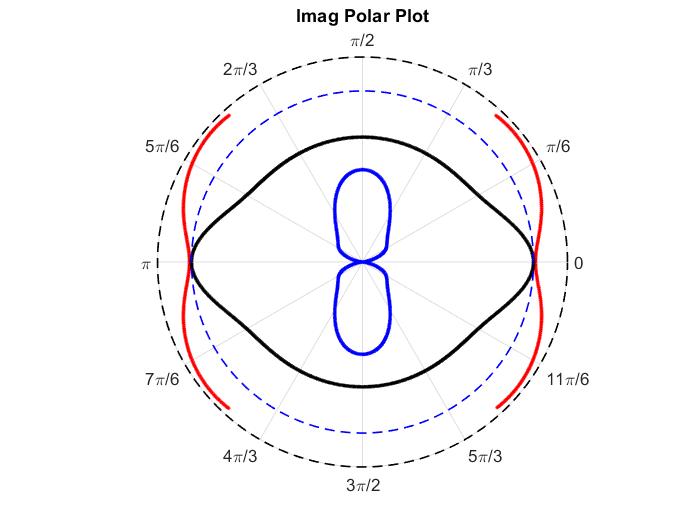}
         \end{subfigure}   
\caption{The same as in Figure \ref{polar}, but here the polar plots have been obtained for the value $k c_{S}/\Omega_{Be}=0.07$, i.e. shorter wavelengths. }
   \label{polar1}
   \end{figure*}

In the right panel of Figure \ref{polar} we plot the absolute value of the damping rate of the three waves, when the propagation direction is, again, covering the full $2\pi$ range. It is clear that not only the propagation speed of waves depends on the propagation angle of waves, but also the damping rate.

First of all, the slow mode associated to the charged particles has the largest damping rate, and its value takes its minimum when these waves propagate along the direction of the magnetic field. The polar plot of imaginary part of the frequency (but also visible on the diagram corresponding to the real part) shows that these modes do not propagate when the angle of propagation is $\pm 3\pi/9=40^{\circ}$ with respect to the perpendicular direction. 

The polar diagram of the damping rate of neutral slow waves (blue curve) also shows a very anisotropic behaviour. These waves propagate along the magnetic field lines with practically no damping, while their damping rate is maximum when they propagate perpendicular to the magnetic field. Among the three possible modes, the slow waves associated with the neutral species has the smallest damping rate.

Finally, fast waves have their largest damping rate when they propagate along the equilibrium magnetic field (where the damping rate is equal to the damping rate of slow waves), and their smallest damping rate is attained when they propagate across the magnetic field.

The same analysis was repeated for another value of $kc_S/\Omega_{Be}$ (here we considered the value of 0.07) to evidence the change in the properties of waves for shorter wavelengths (higher wavenumbers) and the results we obtained for the real and imaginary part of the frequency are shown in Figure \ref{polar1}. The differently coloured curves correspond to the same modes as in Figure \ref{polar}, but here the reference levels (shown by the dashed lines) are at 5.5 and 6.9 MHz for the real part, and 1.33 and 1.6 MHz, respectively. The change in the size of the polar plot corresponding slow waves attached to the charged species is only apparent, instead the frequencies of slow modes associated to neutrals and the fast mode increase, however, we maintain the proportionality. Comparing the two polar plots we obtained for the real part of the frequencies, it is clear that with decreasing the wavelength, the domain where the frequencies of neutral slow waves and fast waves are approximately equal is increased, for the particular values chosen here the two waves have almost identical frequency for propagation angle corresponding to $\pm \pi/6=30^{\circ}$ with respect to the direction of the equilibrium magnetic field. 

The polar plot corresponding to the damping rate of waves shows that, while the curves corresponding to slow waves connected to charges and fast waves do not show significant variation in terms of wavelength (with the reference curves showing the same values as before), the damping rate of slow waves associated to neutrals increases. Naturally, the polar plots shown here are qualitatively similar to the plots we would obtain for the propagation speed.

 Finally, we ought to point out that the model we considered here is a rather simplistic approach to the very complex realistic interaction between particles. Here we considered that only electrons are magnetised, however, given the very tight coupling of electrons and ions, it is likely that ions will also undergo a partial magnetisation. Similarly, given the high collisional rate between the massive particles, neutrals will also be somehow weakly affected by the magnetisation of electrons. In addition, some processes that have been ignored in this paper, like charge exchange and ionization/recombination, may have important effects. 

\section{Conclusions}

Our study deals with the properties of waves in weakly ionised plasmas, characteristic to the lower regions of the solar atmosphere, where the dynamics is determined mainly by the strong collisional coupling between particles. 

First of all, our results showed that since we are dealing with very short wavelength waves, their dynamics is not affected by gravitational stratification.  This conclusion stands only for models where the frequency of waves is comparable with the collisional frequency between particles. In this case the wavelength of waves is much sorter than the gravitational scale-height. For wave frequencies much lower than the collision frequency, i.e. long wavelengths, collisional damping is weak and stratification effects must certainly be more relevant. For example, the study by Soler et al. (2017) showed that the dominance of either stratification effects (reflection) or collisional effects (damping) clearly depends on the wave frequency. It is easy to estimate that for a temperature of 5770 K (for which the isothermal gravitational scale height is 290 km), stratification effects will be important for waves with frequency lower than 0.25 Hz. In the low chromosphere ($T\approx$ 8000 K), stratification will be important for waves with frequencies smaller than 0.21 Hz.

A comparative study of the collisional frequency associated with various species and the gyro-frequency of charged particles revealed the existence of two distinctive regions, where the first region covers the whole photosphere. In the first region the collisional frequency of charged particles are larger than the gyro-frequencies of charged particles, meaning that here the dynamics of particles is not affected by the presence of magnetic fields and the only waves that can propagate are acoustic in nature. Since the collisional frequency of electrons to ions and neutrals are very close to each other, the wave dynamics was described in a full three-fluid framework. These results show that in a multi-fluid description of plasma in the solar photosphere the effects of the magnetic field can be neglected.

Given the very low ionisation degree of the plasma, the neutral acoustic modes propagate undamped, and their amplitude is not affected by the collisions with other particles. In contrast, the acoustic modes associated with ions and electrons propagate only when their wavenumber is greater that a cut-off value, determined by the collisional frequency with neutrals. Their frequency tends to the frequency of neutral acoustic modes when their wavelength decreases. These waves are damped, but the damping rate does not depend on the wavelength of waves. For small wavelengths there is no distinction between various waves, all propagate with the frequency of the neutral species.

The second region corresponds to a frequency regime bounded by the two gyro-frequencies, so electrons become magnetised, while the dynamics of ions are still mainly driven by the collisions with neutrals. Thanks to the different dynamics of the charged particles, an electric current is generated, together with an electric field that opposes the disassociation of charges. In this region the charged particles become strongly coupled (their collisional frequency is much higher than the collisional frequency with neutrals), meaning that charged particles form a single fluid that interacts with neutrals through collisions. 

In such plasmas waves become magnetoacoustic in nature, however the slow waves associated with neutrals retains much of the characteristics of neutrals sounds waves in Region I, however, it propagates with no damping only when the wave propagates mainly along the magnetic field. In addition to the neutral slow waves, ion slow waves can also propagate, however these waves propagate with a cut-off and their frequency becomes fairly quickly independent on the wavelength of waves (non-dispersive). In contrast, the fast waves are non-dipersive only for very large wavelengths and become attached to neutral slow waves. The Friedrich polar diagram show a strong dependence of frequency and damping rate with the propagation angle of waves.

 Our analysis described the nature and properties of waves in different regions of the weakly ionised solar plasma and in each region we have introduced a new set of equations. Probably a more consistent approach would have been to use everywhere the same set of multi-fluid equations in the approximation of weakly ionized plasma as done by, e.g. Jones and Downes (2011) and reduce the corresponding terms for particular settings.

Our analysis considered only the two-dimensional propagation of magnetocoustic waves. It is obvious that a full, three dimensional propagation, would recover an even richer spectrum of waves, however, this (numerical) investigation would be the subject to future studies.  

\section*{Acknowledgements}

AA acknowledges Umm Al-Qura University and Ministry of Education in the Kingdom of Saudi Arabia for their financial support. VF and GV, and IB is grateful to The Royal Society, International Exchanges Scheme, collaboration with Brazil (IES191114) Chile (IES170301) and Ukraine (IES211177). VF and GV are grateful to Science and Technology Facilities Council (STFC) grant ST/V000977/1. VF would like to thank the International Space Science Institute (ISSI) in Bern, Switzerland, for the hospitality provided to the members of the team on `The Nature and Physics of Vortex Flows in Solar Plasmas'. This research has received partial financial support from the European Union’s Horizon 2020 research and innovation program under grant agreement No. 824135 (SOLARNET).
\section*{Data Availability}

The data underlying this article will be shared on reasonable request to the corresponding author.



\bibliography{example} 



\label{lastpage}
\end{document}